\documentstyle[tighten,pra,aps,multicol,epsfig]{revtex}
\begin{document}

\title{Upper Bound on the region of Separable States near the Maximally Mixed State}
\author{P. Deuar\cite{PDemail}, W. J. Munro and K. Nemoto}
\address{Centre for Laser Science,Department of Physics, University of
Queensland,\\ QLD 4072, Brisbane, Australia}
\date{July 7, 1999}
\maketitle
\begin{abstract}
 A lower bound on the amount of noise that must be added to a
GHZ-like entangled state to make it separable (also called the random
robustness) is found using the
transposition condition. The bound is
applicable to arbitrary numbers of subsystems, and dimensions of Hilbert
space, and is shown to be exact for qubits. The new bound is compared
to previous such bounds on this quantity, and found to be stronger in
all cases. It implies that increasing the number of subsystems,
rather than increasing their Hilbert space dimension is a more
effective way of increasing entanglement.
  An explicit decomposition into an ensemble of separable states, when
the state is not entangled,is
given for the case of qubits.
\end{abstract}
\begin{multicols}{2}

\section{Introduction}
\label{INTRO}

A key distinguishing feature of quantum physics from classical physics is the prediction of a
new kind of correlation between physical quantities, called entanglement. 
Quantum entanglement has been often been referred to as the inseparability of composite 
quantum systems. Such an entangled  composite system is said to be
inseparable because it cannot be prepared by manipulating each
subsystem separately, using only measurements and operations local to
one subsystem at a time. 
If a composite quantum mechanical state 
is specified by some density matrix, how can we tell if the system is 
entangled?

Much work has been done studying the particular case of two subsystems, with 
each in a two dimensional Hilbert space (qubits). There is a  good understanding 
of the entanglement for such systems and in fact a criterion, the partial
transposition condition of Peres\cite{Peres:96}, indicates whether the subsystems 
are entangled. This, however, is a necessary and sufficient condition
only when there are two subsystems, one with Hilbert space dimension
$2$, and the other of dimension $2$ or $3$, as was shown by the
Horodeckis\cite{3Horodecki:96}. For more complex systems, it
only determines whether the state contains distillable entanglement,
however there are also some states with bound entanglement, which are
lumped together with the separable states by this criterion. 

Lewenstein {\it et. al.}\cite{Lewenstein:99} have used the Peres condition 
to consider two subsystems, but with each subsystem now in a $N>2$
dimensional Hilbert space.
\.{Z}yczkowski {\it et. al.}\cite{Zyczkowskietal:98} have, among other
results, shown that all the mixed states in a sufficiently small
neighbourhood of the maximally mixed state are separable. They also
gave a bound on  the size of this neighbourhood for small composite
systems. 
Vidal and Tarrach\cite{VidalT:99} gave a lower bound on the size of
this neighbourhood for arbitrary composite states, of any number of
subsystems. Schack and Caves\cite{SchackC:99} gave bounds for
composite systems composed of many qubits ($N=2$).


It has been pointed out by Braunstein {\it
et. al.}\cite{Braunsteinetal:98} that 
Maximally-entangled states of the GHZ type, with noise added, are
connected to recent proposals for NMR quantum computing. They are also
relevant to fundamental tests of quantum mechanics using Bell
inequalities.  
  The separability of such noisy generalised singlet states has been 
considered by various authors recently, mostly for the case of qubits
(subsystems with Hilbert space dimension two). Schack and
Caves\cite{SchackC:99}, using the approach of Braunstein {\it et. al.}
have obtained an exact boundary condition for the Werner
states\cite{Werner:89} (two qubits), while Caves and
Milburn\cite{CavesM:99} extended the approach to two q-trits (Hilbert
space dimension three). The Horodeckis\cite{HorodeckiH:99} gave exact
bounds for the case of two subsystems.

In this paper we extend the approach of Peres\cite{Peres:96} to
 consider such noisy generalised singlet states in the general case of
 $D$ subsystems with each 
subsystem in a $N$ dimensional Hilbert space. A parameter $\epsilon$ 
specifies the amount of the pure GHZ-type maximally-entangled states present compared 
with the maximally-mixed noise state.We then 
ask and answer the following question: what is the maximum value of $\epsilon$  
for this state to be entangled?
 Another fundamental question is also considered. To 
create as entangled a state as possible, is it simply better to 
increase the dimension of the subsystems or is it better to increase 
the number of subsystems?

\section{Generalised Werner states}
\label{STATE}
The states considered here, consist of a mixture of the
maximally-mixed noise
states, and the GHZ-type entangled states, where the number $D$ of subsystems
over which entanglement occurs, and the Hilbert space dimension $N$ of
each subsystem (equal for all $D$ subsystems) can take on any values
$N > 1$, $D>1$.  The relative proportion of GHZ-type states is
controlled by the parameter $\varepsilon$, where $\varepsilon = 1$ corresponds to a
pure GHZ-type state, while $\varepsilon=0$ corresponds to the maximally mixed
state. Explicitly, the state is given by the density operator
\begin{mathletters}
\label{p=}
\begin{eqnarray}
\hat{\rho}\:\: &=& (1-\varepsilon)\hat{\rho}_n + \varepsilon\hat{\rho}_e, \\
\hat{\rho}_n&=& \frac{1}{N^D} \sum_{i_1,i_2,\dots,i_D = 1}^{N} 
\left|{i_1 i_2 \dots i_D}\right\rangle\left\langle{i_1 i_2 \dots i_D}\right|,\\
\hat{\rho}_e&=&  \frac{1}{N}\sum_{n,m = 1}^N \left|{n n \dots 
n}\right\rangle\left\langle{m m\dots m}\right|,
\end{eqnarray}
\end{mathletters}
where 
\begin{eqnarray}
\label{ketdef}
\left|{i_1 i_2 \dots i_D}\right\rangle = \left|{i_1}\right\rangle\otimes\left|{i_2}\right\rangle\otimes \cdots
\otimes \left|{i_D}\right\rangle.
\end{eqnarray}
Here $\left|{i_k}\right\rangle$ represents one of $N$ complete orthogonal basis states for 
 the $k^{\rm th}$ subsystem, while $\hat{\rho}_n$ is the density operator for the
maximally mixed state, and $\hat{\rho}_e$ for the GHZ-like state. 

The Werner state\cite{Werner:89}, consisting of a singlet state and
some noise, is the simplest $D=N=2$ case, hence we call these
generalised Werner states.  The state $\hat{\rho}$ can be viewed as a
generalised singlet state $\hat{\rho}_e$ after it has emerged from a
depolarising channel. 

\section{Separability}
\label{PROBLEM}
The separability of particular cases of the state (\ref{p=}), and of
more general states,  has 
been considered previously by a number of authors
\cite{Zyczkowskietal:98,VidalT:99,SchackC:99,Braunsteinetal:98,Werner:89,CavesM:99,HorodeckiH:99,Muraoetal:98,2Horodecki:96}.
Firstly, it was shown \cite{Muraoetal:98,2Horodecki:96} that for the two-qubit ($N=2,D=2$) case, (The 
Werner state) $\hat{\rho}$ is separable for \mbox{$\varepsilon \le 1/3$}, and entangled otherwise, whereas for three qubits ($N=2,D=3$),
Schack and Caves\cite{SchackC:99} found the state to be separable for 
\mbox{$\varepsilon \le 1/5$}. Soon after, Horodecki and
Horodecki\cite{HorodeckiH:99} found the exact result for arbitrary numbers
of qubits:
\begin{equation}
\varepsilon_{\text{separable}} \le \frac{1}{1+N}.
\end{equation}
  Schack and Caves also found bounds on the size of the separable
neighbourhood around the maximally mixed state for totally general
states of many qubits ($N=2$). For \mbox{$D=4$} and \mbox{$D=5$} these
are \mbox{$\varepsilon\le 1/33$}  and \mbox{$\varepsilon\le 1/243$}
respectively, and for higher $D$ are given by
\begin{equation}
\varepsilon_{\text{separable}} \le \left\{ 
\begin{array}{ll}
1\ /\ (1+2^D+2^{2D-2})  &{\rm{if\;D\;even}} \\	
1\ /\ (1-2^D+2^{2D-2})  &{\rm{if\;D\;odd}} 
\end{array}
\right.
\end{equation}

What about the more general case when $N$ and $D$ are arbitrary? For what values of 
$\varepsilon$ is the state given by 
(\ref{p=}) separable?
Vidal and Tarrach\cite{VidalT:99} gave a maximum bound for the random
robustness $R$ of arbitrary multi-component states. For the
states considered here, the critical value of $\varepsilon$ at which
the states change from being entangled to separable is
\mbox{$\varepsilon_c = 1/(1+R(\hat{\rho}_e||\hat{\rho}_n))$} (in the
notation of ref.\cite{VidalT:99}). So according to that
bound, 
\begin{equation}\label{VTbound}
\varepsilon_{\text{entangled}} > \frac{1}{(1+N/2)^{D-1}}
\end{equation} 

 We will prove in the next section 
(section~\ref{PROOF}) that the states given by
(\ref{p=}), are always entangled if 
\begin{eqnarray}
\label{Eo=}
\varepsilon_{\text{entangled}} > \frac{1}{N^{D-1}+1}.
\end{eqnarray}
That this bound is strong for qubits (\mbox{$N=2$}) is shown in
appendix~(\ref{STRONG}), and an explicit decomposition into product
states is given for the case of separable $\hat{\rho}$.

\section{Outline of the proof of (6)}
\label{PROOF} 
Peres\cite{Peres:96} has shown that a necessary condition for a state consisting of two subsystems to be
separable is that the partial transpose  of the density matrix over 
one  of the subsystems, and the partial transpose over the other
subsystem,  have positive eigenvalues.  However, this is a necessary and sufficient condition only
when one of the subsystems has Hilbert space dimension $2$ or $3$,
and the other dimension $2$, as was shown by the
Horodeckis\cite{3Horodecki:96}. That paper went on to give a necessary
and sufficient condition for a state to be separable. Nevertheless the Peres condition is
just what is needed for an upper bound on $\varepsilon$ for separable
states. i.e. any states which break the condition are entangled,
although some which satisfy it may also be, but do not have to be,
entangled. 
  It is worth noting that any states which satisfy the Peres condition
but are entangled are said to contain only ``bound'' entanglement, as
it cannot be used for teleportation, nor distilled by the process of
entanglement distillation.

Firstly, note that the Peres condition is easily extended to more than
two entangled subsystems. If there are $D$ subsystems, one simply
chooses some group of \mbox{$M < D$} original subsystems to be called
half-system number $1$, and the remaining subsystems to be called
half-system number $2$. If for any such group of subsystems, an
eigenvalue of the partial transpose of $\hat{\rho}$ over half-system
number $1$ (say) is negative, then $\hat{\rho}$ is entangled. Thus to
use the Peres condition to full advantage, one must consider all such
groups of subsystems. 
  
The state (\ref{p=}) is convenient in this respect, because it is
unchanged under relabeling of the subsystems (evident by
inspection). Thus the eigenvalues of $\rho^T_\alpha$, the partial
transpose of $\hat{\rho}$ over the set $\alpha$ of subsystems, need
only be looked at for $D/2$ (rounded
down) sets of subsystems to extract the maximum benefit from the Peres
condition. In particular, a choice of sets of subsystems can be
$\alpha_M = \{1,2,\dots,M\}$ where $M = 1,2,\dots,D/2$,
and $\alpha_M$ contains the labels of the subsystems to be considered
as members of  the $M$th half-system.

Firstly let's consider $M=1$, i.e. the first subsystem's entanglement
with the remaining $D-1$ of them.
  The partially transposed density matrix is 
\begin{mathletters}
\label{pt1=}
\begin{eqnarray}
\rho^T_1 &=& (1-\varepsilon){\rho}^T_n + \varepsilon{\rho}^T_e, \\
{\rho}^T_n&=& \frac{1}{N^D} \sum_{i_1,i_2,\dots,i_D = 1}^{N} 
\left|{i_1 i_2 \dots i_D}\right\rangle\left\langle{i_1 i_2 \dots i_D}\right|,\\
{\rho}^T_e&=&  \frac{1}{N}\sum_{n,m = 1}^N \left|{m n \dots n}\right\rangle\left\langle{n m\dots m}\right|. 
\end{eqnarray}
\end{mathletters}

Since all of the elements of $\hat{\rho}$, (hence $\rho^T_1$) are
finite, the eigenvalues of $\rho^T_1$ must also be finite. Now,
exploiting the general continuity property of eigenvalues of
$\hat{\rho}_1^T$, we conclude that 
 at the
value of \mbox{$\varepsilon=\varepsilon_o$} above which the Peres condition indicates the state is
entangled, one or more eigenvalues of $\rho^T_1$ must be
zero, since they are all positive for \mbox{$\varepsilon <
\varepsilon_o$}, and at least one is negative for \mbox{$\varepsilon >
\varepsilon_o$}. i.e. for some nonzero eigenvector 
\begin{equation}
\label{eigenvector}
\left|{\psi}\right\rangle = \sum_{j_1,j_2,\dots,j_D = 1}^N \psi_{j_1 j_2 \dots j_D}
\left|{j_1 j_2 \cdots j_D}\right\rangle  \neq 0
\end{equation}
we must have $\rho^T_1(\varepsilon_o) \left|{\psi}\right\rangle = 0$. Expanded, this gives
\begin{eqnarray}
\label{ptpsi}
\frac{1-\varepsilon_o}{N^D}\sum_{i_1,i_2,\dots,i_D = 1}^N &\psi_{i_1 i_2 \dots
i_D}&\left|{i_1 i_2 \cdots i_D}\right\rangle \nonumber \\
+ \frac{\varepsilon_o}{N}\sum_{n,m = 1}^N &\psi_{n
m \dots m}&\left|{m n \cdots n}\right\rangle = 0.
\end{eqnarray}
Equation (\ref{ptpsi}) can explicitly be written out as $N^D$ equations 
\begin{eqnarray}
\label{eigeqns}
&(1-\varepsilon_o)&\psi_{i_1 i_2 \dots i_D} \nonumber \\ 
&+& \varepsilon_o N^{D-1}\delta_{i_2 i_3}\delta_{i_2
i_4}\cdots\delta_{i_2 i_D} \psi_{i_2 i_1 \dots i_1} = 0,
\end{eqnarray}
where $\delta_{ab} = 1$ if $a=b$, $0$ otherwise. Now if one or more of the
$i_a : a=3,\dots,D$ does not equal $i_2$, then that equation is
satisfied only if $\varepsilon = 1$ or $\psi_{i_1 i_2 \dots i_D}= 0$. The first
case is not of interest here, as $\varepsilon = 1$ corresponds to our 
maximally entangled GHZ like states, so we choose $\psi_{i_1,i_2 \dots i_D} = 0$. 

The rest of the equations where  $i_2 = i_3 = \cdots = i_D$, separate
into $ N(N-1)/2$ coupled sets of two equations of the identical form
\begin{eqnarray}
\label{eigeqn}
(1-\varepsilon_o)\psi_{a b \dots b}  + \varepsilon_o N^{D-1} \psi_{b a \dots a} &=& 0, \\\label{eigeqn2}
(1-\varepsilon_o)\psi_{b a \dots a}  + \varepsilon_o N^{D-1} \psi_{a b \dots b} &=& 0. 
\end{eqnarray}
These have solutions if $\psi_{a b \dots b} = \psi_{b a \dots a} = 0$,
but this would imply $\left|{\phi}\right\rangle = 0$, which was specifically excluded
in (\ref{eigenvector}). Otherwise, these coupled two equations are only satisfied if
\begin{eqnarray}
\label{Eo}
\varepsilon_o = \frac{1}{N^{D-1}+1}.
\end{eqnarray}
So for $\left|{\psi}\right\rangle \neq 0$, at least one such coupled set of two
equations leads to the expression (\ref{Eo}). This is the only
candidate for the point where the Peres condition becomes satisfied. 

Now it can be easily seen that in the total-noise
case $\varepsilon = 0$, $\rho^T_1 = \hat{\rho}$ and all the eigenvalues of
$\rho^T_1$ are $1/(N^D)$. In the no-noise case ($\varepsilon = 1$),
proceeding in similar fashion to before, the
eigenvalues $\lambda$ of $\rho^T_1$ must satisfy 
\begin{eqnarray}
\label{E=1ptpsi}
\frac{1}{N}\sum_{n,m = 1}^N &\psi_{n
m \dots m}&\left|{m n \cdots n}\right\rangle \nonumber \\
=& \lambda&\sum_{i_1,i_2,\dots,i_D=1}^N
\psi_{i_1 i_2 \dots i_D}\left|{i_1 i_2 \cdots i_D}\right\rangle.
\end{eqnarray}
This gives $\psi_{i_1 i_2 \dots i_D} = 0$ if for some
$a=3,4,\dots,D$, $i_2 \neq i_a$, and leads to sets of two coupled equations of
the form
\begin{eqnarray}
\label{E1eigeqn}
 \psi_{a b \dots b} &=& \lambda N \psi_{b a \dots a}, \\
 \psi_{b a \dots a} &=& \lambda N \psi_{a b \dots b}.
\end{eqnarray}
These have the solutions $\psi_{a b \dots b} = \psi_{b a \dots a} = 0$
or 
\begin{eqnarray}\label{specialcase}
\lambda = \pm\frac{1}{N^2}.
\end{eqnarray}
As before, for nonzero eigenvectors,
the first cannot be true. So, finally, for the no-noise state, at
least one eigenvalue of $\rho^T_1$ must be negative, and equal to 
\begin{eqnarray}
\lambda = -\frac{1}{N^2}.
\end{eqnarray}
Thus, finally, since $\varepsilon_o$ is the only value of $\varepsilon$ where an eigenvalue of
$\rho^T_1$ is zero, all eigenvalues are positive for $\varepsilon = 0$, and
an eigenvalue is negative for $\varepsilon = 1$, there must be at least one
negative eigenvalue for $\varepsilon > \varepsilon_o$ given by (\ref{Eo}). 

When one proceeds in the same fashion when $M = 2, 3, \dots , D/2$,
one always gets coupled sets of two equations identical in form to
(\ref{eigeqn}), as can be seen by inspection,  so nothing new is
found. Thus the result of section~(\ref{STATE}) is indicated. 

\section{Comparison to known bounds}
\label{COMPARISON}

\begin{figure}[htb]
\epsfig{figure=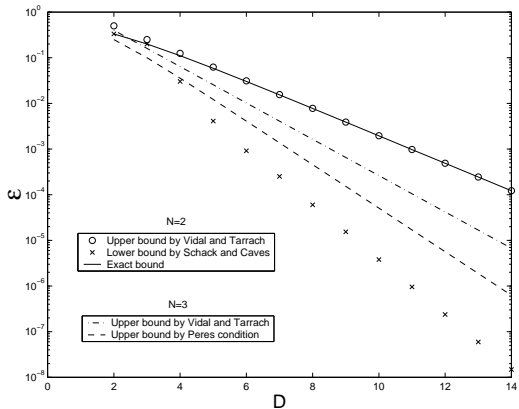,width=70mm}
\caption{\label{N=2} Bounds on the value of $\varepsilon$ for which
the states (\ref{p=}) become separable or entangled, for the qubit ($N=2$) and
q-trit ($N=3$) cases, shown on the same plot. When $\varepsilon$ is above the upper bounds or
the exact bound,
$\hat{\rho}$ is entangled, and when $\varepsilon$ is below the lower
bound or the exact bound, $\hat{\rho}$ is always separable. When
$\varepsilon$ is below an upper bound, $\hat{\rho}$ may be separable
or bound entangled.}
\end{figure}

For the qubit case ($N=2$), as seen in figure~(\ref{N=2}), equation
(\ref{Eo=}) gives an exact bound on the values of $\varepsilon$ that
divide 
separable from  entangled states of the form (\ref{p=}).  
One sees
that the upper bound\ (\ref{VTbound}) derived from the work of Vidal and Tarrach comes very close
to the exact value for the qubit case. 
This exact bound is also 
greater than 
those lower bounds previously found by Schack and
Caves\cite{SchackC:99}, as expected.

For the two-subsystem ($D=2$) case, the bound (\ref{Eo=}) 
agrees with the exact one found by the Horodeckis\cite{HorodeckiH:99}.

For other values of $D$ and $N$, the results on random robustness by
Vidal and Tarrach, lead to an upper bound which is considerably
weaker than the upper bound\ (\ref{Eo=}) given by the partial
transposition condition.  The upper bound found here, actually gives a
stronger bound on the random robustness of entanglement of states given by\ (\ref{p=}).
\begin{equation}
R(\hat{\rho}_e||\hat{\rho}_n)  \le N^{D-1}
\end{equation}

It is interesting to note that in the border cases when $N=2$ or
$D=2$, the bound\ (\ref{Eo=})  is in fact an exact bound. 
This is  despite the fact
 that the Peres condition does not
necessarily give a strong bound for such states. This leads one
to the tentative conjecture that for noisy GHZ-type states of the form
(\ref{p=}), the Peres condition may in general give a strong upper
bound.

\begin{figure}[htb]
\epsfig{figure=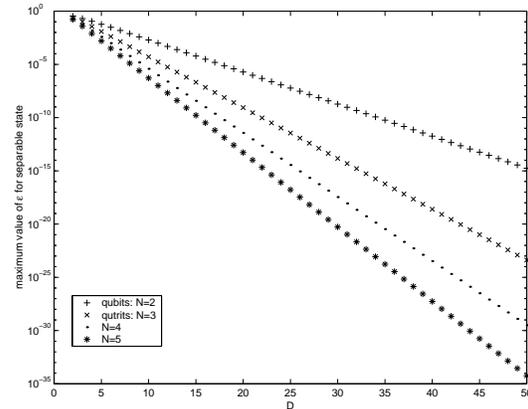,width=70mm}
\caption{\label{varD} Upper bounds on $\varepsilon$ for separable states of
the form (\ref{p=}) on a logarithmic scale. Variation with subsystem
Hilbert-space dimension $N$ is shown.} 
\end{figure}

Looking at figure~(\ref{varD}), one sees that as the Hilbert space
dimension of the subsystems increases, the upper bound on $\varepsilon$
rapidly decreases, indicating that the entanglement becomes
stronger. 

As the bound (\ref{Eo=}) is completely general in $N$ and $D$, it
does provide some answers to the question of what raises entanglement
more: creating more entangled subsystems, or increasing their dimension?
Since the bound on $\varepsilon$ decreases exponentially
with $D$, but only polynomially with $N$, one concludes that
increasing the number of subsystems is a more effective way of
increasing the entanglement.

\section{Conclusion}
\label{CONCLUSION}

The results presented here based on the Peres condition provide a
lower bound on the parameters $\varepsilon$ in\ (\ref{p=}) above which
the generalised Werner states are always entangled. Furthermore, it
gives an exact bound on this parameter for the case of many qubits.
An explicit simple expression is derived that depends 
on $D$, the number of subsystems over which entanglement occurs, and $N$ the Hilbert space dimension  
of each subsystem. Apart from the few cases ($D=2$;$N=2$ and $D=3$)
where this bound was known exactly previously, the new bounds are
stronger than previously known ones. 

This work also sheds light on the question of whether to 
increase quantum entanglement in a system, is it better to 
create more entangled subsystems, or to increase the dimension of the 
existing subsystem. As the bound on $\varepsilon$ decreases exponentially
with $N$, but only polynomially with $D$, increasing the number of subsystems 
is a much more effective way of increasing entanglement. 

To conclude, the Peres partial transposition condition has provided a 
good upper bound for determining the separability of a generalised $N$, $D$ Werner
state. However for other systems it is known that this transposition condition 
fails to give strong results, thus when this condition is useful, and
when not, remains an interesting question.

\acknowledgments{We are grateful to Gerard Milburn for discussion 
about entanglement and separability. W.J.M acknowledges the support of the Australian Research Council.}

\appendix
\section{Proof that (6) is exact for qubits}
\label{STRONG}
  We wish to show that the bound\ (\ref{Eo=}) is exact for qubits
($N=2$), and to find the product states which combine to give the
state when it is separable.

Starting off similarly to Schack and Caves\cite{SchackC:99},
the $D$-subsystem generalisation of the Werner state ((\ref{p=}) with $N=2$)
can be written in terms of Pauli matrices:
\begin{mathletters}\label{rho}
\begin{equation}
  \hat{\rho}(\varepsilon) = \frac{1}{2^D}\left\{ (1-\varepsilon)I^{\otimes D} 
         + \frac{\varepsilon}{2}\hat{E}\right\},
\end{equation}
where
\begin{eqnarray}
  \hat{E} =& (I + \sigma_3)^{\otimes D} 
     + (I - \sigma_3)^{\otimes D}\nonumber\\
 &+ (\sigma_1 + i\sigma_2)^{\otimes D}
     + (\sigma_1 - i\sigma_2)^{\otimes D},
\end{eqnarray}
\end{mathletters}
and $\sigma_i$ are the Pauli matrices, $I$ is the two-dimensional
identity matrix, and for conciseness, the following notation is used:
\begin{equation}
  (A)^{\otimes D} = (A) \otimes \cdots \otimes (A) \qquad \text{$D$ times}.
\end{equation} 

To show that (\ref{Eo=})  is actually a strong bound, is suffices to
find an expansion of $\rho$ in terms of a positive sum of direct tensor products of
density matrices for this value of 
\begin{equation}\label{ec}
\varepsilon = \varepsilon_c = \frac{1}{2^{D-1}+1}.
\end{equation}

By analogy with the results of Schack and Caves \cite{SchackC:99}, it
has been  guessed that an expansion  of $\rho(\varepsilon_c)$ in terms of
the density matrices
\begin{equation}\label{P}
  P_{\pm i} = \frac{1}{2}(I \pm \sigma_i ) \qquad \text{ ($i = 1,2,3$)},
\end{equation}
is given by the following: 
\begin{equation}\label{expguess}
\hat{\rho}_g = \frac{\varepsilon_c}{2}\left( P_3^{\otimes D} + P_{-3}^{\otimes D} \right) 
  + \frac{\varepsilon_c}{2^{D-1}} \sum P_{x_1} \otimes P_{x_2} \otimes
  \cdots \otimes P_{x_D}.
\end{equation}
Here the sum is over all permutations of $D$  indices $x_1,x_2,\dots,x_D$
satisfying the following conditions:
\begin{enumerate}
\item   $x_i \in \{ 1, -1, 2, -2 \}$.
\item  The number of $x_i \in \{ 2, -2 \}$ is even (or zero).
\item If the number of $x_i \in \{ 2, -2\}$ is a multiple of four
(or is  zero),
then the number of \mbox{$x_i \in \{ -1, -2 \}$} is even (or zero).
\item If the number of $x_i \in \{ 2, -2\}$ is not a multiple of four,
then the number of \mbox{$x_i \in \{ -1, -2 \}$} is odd.
\end{enumerate}

The proof of this proceeds by starting with the above guess, and showing
that it is a correct one. As it turns out, the hard work is in
actually writing down the guess mathematically, so let us begin with this.

To begin with, note that 
\begin{eqnarray}\label{T0}
{\cal T}_0 &=& (P_1 + P_2 + P_{-1} + P_{-2})^{\otimes D}\nonumber\\  &=& \sum P_{x_1}
\otimes P_{x_2} \otimes \cdots \otimes P_{x_D},
\end{eqnarray}
where the sum is over \emph{all} permutations of $D$ indices
$x_1,x_2,\dots,x_N \in \{ 1, 2, -1, -2 \}$. 
Also note that 
\begin{equation}\label{S0}
{\cal S}_0 = (P_1 + P_2 - P_{-1} - P_{-2})^{\otimes D}
\end{equation}
will give a sum over all these permutations, except that all terms
which have an odd number of indices in $\{-1, -2\}$ will be subtracted
rather than added like in ${\cal T}_0$. Thus one can see that 
\begin{mathletters}
\label{R}
\begin{equation}
{\cal R}_0^e = \frac{1}{2}[{\cal T}_0 + {\cal S}_0]
\end{equation}
 will give a sum over 
permutations of $D$ indices, like ${\cal T}_0$, except that only terms
where
the number of indices in
$\{-1,-2\}$ is even will be included. Similarly 
\begin{equation}
{\cal R}_0^o = \frac{1}{2}[{\cal T}_0 - {\cal S}_0]
\end{equation}
\end{mathletters}
will give a sum over only those terms in which the 
number of indices in $\{-1, -2\}$ is odd. 

Now consider some more similar expressions.
\begin{equation}\label{T1}
{\cal T}_1 = (P_1 - P_2 + P_{-1} - P_{-2})^{\otimes D}.
\end{equation}
${\cal T}_1$ will give a sum over all index permutations, except that
all terms which have an odd number of indices in $\{2, -2\}$ will be
subtracted rather than added like for ${\cal T}_0$.
\begin{equation}
{\cal T}_2 = (P_1 + iP_2 + P_{-1} + iP_{-2})^{\otimes D}.
\end{equation}
${\cal T}_2$ will give a similar sum over all permutations, but terms
in which the number of indices in $\{2,-2\}$ is a multiple of
four (or is zero) will be added, terms in which this number is even, but not a
multiple of four,  will be subtracted, terms in which this
number is one more than a multiple of four will be added and
multiplied by $i$, and terms for which this number is one less than a
multiple of four will be subtracted and multiplied by $i$. 
\begin{equation}
{\cal T}_3 = (P_1 - iP_2 + P_{-1} - iP_{-2})^{\otimes D}.
\end{equation}
${\cal T}_3$ is the complex conjugate of ${\cal T}_2$. 
It can be seen (after a little thought) that $\frac{1}{4}[{\cal T}_0 +
{\cal T}_1 + {\cal T}_2 + {\cal T}_3]$ will give a sum over only those
terms in which the number of indices in $\{2, -2\}$ is a multiple of
four (or is zero). Similarly, $\frac{1}{4}[{\cal T}_0 +
{\cal T}_1 - {\cal T}_2 - {\cal T}_3]$ will give a sum over only those
terms in which the number of indices in $\{2, -2\}$ is even, but not a
multiple of four.

Analogously to equation\ (\ref{S0}) define
\begin{mathletters}\label{Si}\begin{eqnarray}
{\cal S}_1 = (P_1 - P_2 - P_{-1} + P_{-2})^{\otimes D}, \\
{\cal S}_2 = (P_1 + iP_2 - P_{-1} - iP_{-2})^{\otimes D}, \\
{\cal S}_3 = (P_1 -i P_2 - P_{-1} + iP_{-2})^{\otimes D}, 
\end{eqnarray}\end{mathletters}
and one can define expressions for ${\cal R}_i^e$ and ${\cal R}_i^o$
for $i = 0,1,2,3$ analogously to equations\ (\ref{R}).
So following the same reasoning as previously, the sum of all terms in which the number of indices in $\{2, -2\}$
is a multiple of four, and the number of indices in $\{-1, -2\}$ is
even is given by 
\begin{equation}
\frac{1}{4}[ {\cal R}_0^e +{\cal R}_1^e +{\cal R}_2^e +{\cal R}_3^e ].
\end{equation}

And thus finally, the sum in the  second term of the guess $\rho_g$  (
equation\ (\ref{expguess}) ) can be
written 
\begin{eqnarray}
 \frac{1}{4}[  {\cal R}_0^e +{\cal R}_1^e +{\cal R}_2^e +{\cal R}_3^e]
 + \frac{1}{4}[  {\cal R}_0^o +{\cal R}_1^o -{\cal R}_2^o -{\cal
 R}_3^o]\nonumber \\
= \frac{1}{4}[ {\cal T}_0 + {\cal T}_1 + {\cal S}_2 + {\cal S}_3 ].
\end{eqnarray}
So the guess that has been made (i.e. equation\ (\ref{expguess})) can
be rewritten 
\begin{equation}\label{Pguess}
\rho_g =  \frac{\varepsilon_c}{2}\left( P_3^{\otimes N} + P_{-3}^{\otimes N} \right) 
  + \frac{\varepsilon_c}{2^{N+1}} \left[ {\cal T}_0 + {\cal T}_1 + {\cal
  S}_2 + {\cal S}_3 \right], 
\end{equation}
which is explicitly (via the expressions\ (\ref{T0}), (\ref{S0}),
(\ref{T1}), (\ref{Si}))  a positive sum of direct tensor products of density
matrices, and thus is separable. The only question that remains is
whether $\rho_g = \rho(\varepsilon_c)$ ?

This is the easy part. It is seen using the  expression\ (\ref{P})
that
\begin{mathletters}\begin{eqnarray}
 {\cal T}_0 &=& (2I)^{\otimes N} = 2^N I^{\otimes N}, \\
 {\cal T}_1 &=& 0, \\
 {\cal S}_2 &=& (\sigma_1 + i\sigma_2)^{\otimes N}, \\
 {\cal S}_3 &=& (\sigma_1 - i\sigma_2)^{\otimes N}, \\
 P_3^{\otimes N} &=& 2^{-N} (I + \sigma_3)^{\otimes N}, \\
 P_{-3}^{\otimes N} &=& 2^{-N} (I - \sigma_3)^{\otimes N}, 
\end{eqnarray}\end{mathletters}
so 
\begin{equation}
\rho_g = 
      \frac{\varepsilon_c}{2} I^{\otimes N}
+ \frac{\varepsilon_c}{2^{N+1}}\hat{E},
\end{equation}
which only seemingly differs from equation\ (\ref{rho}) by the first
term, but using the expression for $\varepsilon_c$ (\ref{ec}), one finds
that these first terms are equal also.
\begin{equation}
\frac{1-\varepsilon_c}{2^N} = \frac{\varepsilon_c}{2},
\end{equation}
so $\rho_g = \rho$, the guess was correct, and thus the bound
$\varepsilon_c$ is strong.

${\cal QED}$


\end{multicols}

\end{document}